\documentclass[aps,prl,preprint,groupedaddress,showpacs]{revtex4-1}

\usepackage{amsmath}
\usepackage{amssymb}
\usepackage{graphicx}
\usepackage{dcolumn}
\usepackage{bm}
\usepackage{epstopdf}

\begin{document}


\title{Coexistence of bulk and surface polaritons in a~magnetic-semiconductor superlattice influenced by a~transverse magnetic field}


\author{Vladimir~R.~Tuz}
\email[]{tvr@rian.kharkov.ua}
\affiliation{International Center of Future Science, Jilin University, 2699 Qianjin Street, Changchun 130012, China} 
\altaffiliation{State Key Laboratory on Integrated Optoelectronics, College of Electronic Science and Engineering, Jilin University, 2699 Qianjin Street, Changchun 130012, China} 
\altaffiliation{Institute of Radio Astronomy of
National Academy of Sciences of Ukraine, 4, Mystetstv Street, Kharkiv 61002, Ukraine} 

\author{Volodymyr~I.~Fesenko}
\affiliation{Institute of Radio Astronomy of
National Academy of Sciences of Ukraine, 4, Mystetstv Street, Kharkiv 61002, Ukraine} 

\author{Illia~V.~Fedorin}
\affiliation{National Technical
University `Kharkiv Polytechnical Institute', 21, Polytechnichna Street, Kharkiv 61002, Ukraine}

\author{Hong-Bo~Sun}
\affiliation{State Key Laboratory on Integrated Optoelectronics, College of Electronic Science and Engineering, Jilin University, 2699 Qianjin Street, Changchun 130012, China} 

\author{Wei~Han}
\affiliation{College of Physics,
Jilin University, 2699 Qianjin Street, Changchun 130012,  China}
\altaffiliation{International Center of Future Science, Jilin University, 2699 Qianjin Street, Changchun 130012, China} 


\date{\today}

\begin{abstract}
It is demonstrated that the effect of coexistence of bulk and surface polaritons within the same frequency band and wavevector space can be achieved in a magnetic-semiconductor superlattice providing a conscious choice of characteristic resonant frequencies and material fractions of the structure's underlying components as well as geometry of the external static magnetic field. The study is based on the effective medium theory which is involved to calculate dispersion characteristics of the long-wavelength electromagnetic modes of ordinary and extraordinary bulk polaritons and hybrid EH and HE surface polaritons derived via averaged expressions with respect to the effective constitutive parameters of the superlattice.
\end{abstract}


\maketitle

\section{\label{sec:intro}I. Introduction}

In addition to traditional plasmonic systems in which the presence of a metal-dielectric interface is implied, heterostructures capable to support a combined plasmon and magnetic functionality are of great interest. This interest is twofold. First, a number of magneto-optical effects can be greatly increased in such artificial systems due to the electromagnetic field enhancement associated with the plasmon-polariton resonance. Second, which, in fact, is a subject of interest in this paper, by providing a specific structure's design with a conscious choice of its underlying constitutive components, there appears a possibility to modify the plasmon dispersion features by utilizing an external magnetic field as a driving agent in order to realize some modulation and tuning mechanisms. It opens a prospect towards active tunable plasmonic devices, and, in particular, such structures have already found a number of practical applications in the fields of gas- and bio-sensors, and in integrated photonic devices for telecommunications \cite{Armelles_AdvOpticalMater_2013}.

From the viewpoint of theoretical physics, presence of the combined plasmon and magnetic functionality involves consideration of problems related to certain collective excitation (like phonons, plasmons, magnons, etc. \cite{Kaganov_PhysUsp_1997}), which can appear in various magneto-optically active heterostructures. Nevertheless, these different types of excitation can be treated within the overall concept of polaritons \cite{Mills_RepProgPhys_1974}. In the framework of this concept polaritons are considered as modes of the electromagnetic field, and their description is fulfilled on the basis of macroscopic Maxwell's equations, where polaritons are considered as modes existing in a bulk material (\textit{bulk polaritons}) as well as on a medium surface (\textit{surface polaritons}). Therefore, the electromagnetic features of polaritons are closely related to the constitutive properties of a medium, and, in particular, to the resonant states in the frequency dependence of its macroscopic dielectric and magnetic functions (e.g. permittivity and permeability).

Applying such an approach properties of polaritons in heterostructures influenced by an external static magnetic field have been studied by many authors \cite{Burstein_PhysRevLett_1972, Burstein_JPhysC_1973, Burstein_PhysRevB_1974, Camley_PhysRevB_1982, Elmzughi_PhysRevB_1995, Wang_PhysRevB_1995, kushwaha_2001_plasmons, Rapoport_NewJPhys_2005, Tagiyeva_JPCS_2007, Tagiyeva_J_Supercond_Nov_Magn_2012, Fedorin_EurPhysJApplPhys_2014, Baibak_PIERM_2013}. In these works the problem is usually solved within two distinct considerations of \textit{gyroelectric} media (e.g. semiconductors) with magneto-plasmons and \textit{gyromagnetic} media (e.g. ferromagnets) with magnons which involve the medium characterization with either permittivity or permeability tensor having asymmetric off-diagonal components. This distinction is convenient due to the various physical mechanisms which cause the corresponding resonant state to manifest itself in different parts of spectrum. Indeed, characteristic frequencies of permittivity are mostly confined to the optical range, whereas those of permeability usually are in the microwave range.

Although characteristic frequencies of dielectric and magnetic functions normally lie far from each other, it is possible to find exceptions to this rule. In particular, a \textit{gyroelectromagnetic} media in which both permeability and permittivity simultaneously are tensor quantities can be implemented artificially by properly combining together gyroelectric and gyromagnetic materials. As a relevant example magnetic-semiconductor heterostructures \cite{Datta_SuperMicro_1985, Munekata_ApplPhysLett_1993, Koshihara_PhysRevLett_1997, Kussow_PhysRevB_2008, Wang_JApplPhys_2000, Ta_JApplPhys_2010, Ta_PhotNanFundApp_2012, Anderson_JApplPhys_2013} can be mentioned that are able to exhibit a gyroelectromagnetic effect from gigahertz up to tens of terahertz \cite{Jungwirth_RevModPhys_2006}. It should be noted, that in recent years such composites are usually considered within the theory of metamaterials, in the framework of which they are discussed from the viewpoint of achieving negative refraction and backward wave propagation \cite{Tarkhanyan_JMMM_2010, Tarkhanyan_PhysicaB_2010}. On the other hand, already derived solutions of problems with respect to polaritons demonstrate that in such composite media the electromagnetic field structure appears to be rather complicated.

Indeed, in an unbounded isotropic medium which is characterized by a scalar dielectric or magnetic function there is only one TEM eigenwave (i.e. the normal wave or alternatively the bulk wave), whereas in a bounded medium the wave splits into two transverse waves, namely TM-modes and TE-modes for which the field components appear as a superposition of partial solutions of the wave equation \cite{Ivanov_NATO_2010}. Remarkably, in the context of surface polaritons these transverse modes exist only in the frequency bands, where the dielectric or magnetic functions of two patterning materials have different sign. In fact, the TM-modes can propagate only along the surface of a dielectric (nonmagnetic) medium, whereas on the surface of a magnetic medium the TE-modes can exist.

In gyrotropic media the nature of waves is completely different. In any kind of an unbounded gyrotropic medium (i.e. it can be a gyroelectric medium described by permittivity tensor, a gyromagnetic medium described by permeability tensor as well as a gyroelectromagnetic medium described by both permittivity and permeability tensors) there are two distinct eigenwaves (known as ordinary and extraordinary waves \cite{ginzburg_book_1961}), whereas the surface waves split apart only for some particular configurations (e.g. for the Voigt geometry) and generally they have all six field components. Such waves are classified as hybrid EH-modes and HE-modes \cite{Ivanov_NATO_2010}, and these modes appear as some superposition of longitudinal and transverse waves.

Such a diversity in the electromagnetic field characteristics evidently results in the fact that gyrotropic (and especially gyroelectromagnetic) media exhibit an enormous variety of optical properties. Among them in this paper we focus on the particular effect which is related to the ability of surface polaritons to propagate within the bulk polaritons continua. As was already mentioned, surface polaritons can only exist at the interface between two media having opposite sign in their dielectric or magnetic functions (i.e. their excitation frequency is below the characteristic resonant frequency of one patterning material), whereas bulk polaritons only propagate in the medium with positive both dielectric and magnetic functions (i.e. they exist in the band that exceeds the corresponding characteristic resonant frequencies). Therefore, it is believed that surface polaritons can be only found within the bulk polaritons stopbands. Nevertheless, as we demonstrate in this paper in a gyroelectromagnetic medium this rule can be violated.

We should note, for a magnetic-semiconductor superlattice being in the Voigt geometry such an effect has been already reported in our previous publication \cite{Fesenko_OptLett_2016}. Independently, this effect is also found in a waveguide semiconductor-insulator-semiconductor system with the Voigt configuration of magnetization \cite{Zhu_EurophysLett_2016}. Unlike the previous publications, in this paper, for the first time to the best of our knowledge, we discuss in detail a manifestation of the mentioned effect in a magnetic-semiconductor superlattice being in the polar geometry with respect to propagation of the hybrid waves.

The rest of this paper is organized as follows. In Section~II, we formulate the problem related to bulk and surface polaritons propagating through a magnetic-semiconductor superlattice and describe its solution. In Section~III we discuss peculiarities of dispersion features of ordinary and extraordinary bulk polaritons (Subsection~1) and reveal conditions at which the dispersion curves of surface polaritons can merge inside the areas of existence of extraordinary bulk polaritons (Subsection~2). Finally, Section~IV summarizes the paper. Appendices A and B are given at the end of the paper in order to provide insight into the effective medium theory and the constitutive parameters description used here. In Appendix~C the procedure of deriving dispersion equations of both bulk and surface polaritons is presented in detail for the case of a gyroelectromagnetic medium.

\section{\label{sec:problem}II. Superlattice description and dispersion relations}

\begin{figure}
\centerline{\includegraphics[width=12.0cm]{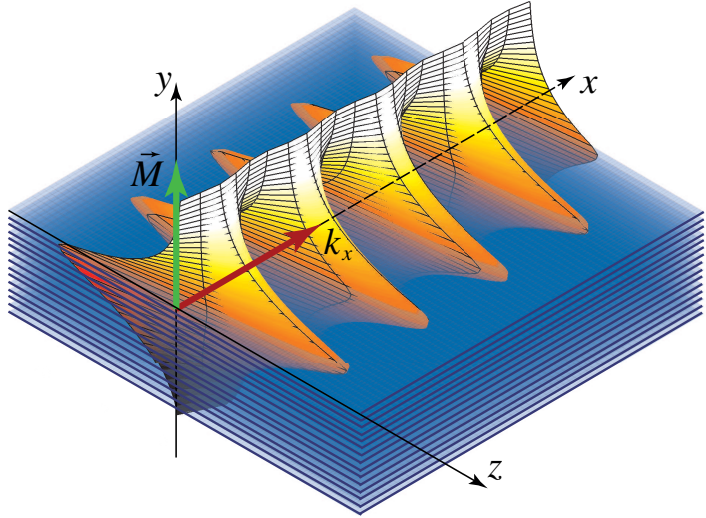}}
\caption{(Color online) The problem sketch and a visual representation of magnitude of the tangential electric field component ($E_z$) of the surface polariton propagating along the $x$-axis on the interface of a superlattice. The superlattice is made from alternating magnetic ($\hat \mu_m$, $\varepsilon_m$) and semiconductor ($\mu_s$, $\hat \varepsilon_s$) layers having thicknesses $d_m$ and $d_s$, respectively. The structure's period is $L$. An external static magnetic field $\vec M$ is applied along the $y$-axis (the polar geometry).}
\label{fig:fig_Struct}
\end{figure}

Thereby, further in this paper we study dispersion features of both bulk and surface polaritons propagating through a superlattice which is composed of a \textit{semi-infinite} stack of identical composite double-layered slabs arranged in the $y$-axis direction (Fig.~\ref{fig:fig_Struct}). Each composite slab within the superlattice consists of magnetic (with constitutive parameters $\hat \mu_m$, $\varepsilon_m$) and semiconductor (with constitutive parameters $\mu_s$, $\hat \varepsilon_s$) layers having thicknesses $d_m$ and $d_s$, respectively. The stack possesses a periodic structure (with period $L = d_m + d_s$) that fills half-space $y<0$ and adjoins an isotropic medium (with constitutive parameters $\mu_0$, $\varepsilon_0$) occupying half-space $y>0$. Therefore, the superlattice interface lies in the $x$-$z$ plane, and along these directions the system is considered to be infinite.

The superlattice is influenced by an external static magnetic field $\vec M$ which is aligned perpendicular to the sample plane, i.e. along the $y$-axis. It is supposed that the strength of this field is high enough to form a homogeneous saturated state of magnetic as well as semiconductor layers, and it is evident, that in the context of polaritons the problem acquires a cylindrical symmetry about the external magnetic field (i.e. it is the polar geometry). Nevertheless, for certainty, we consider that electromagnetic waves propagate along the $x$-axis, therefore, the wavevector $\vec k$ of the surface polaritons has components $\{k_x,\pm i\kappa,0\}$, where $\kappa$ is responsible for the wave attenuating away from the interface, i.e. in the positive ($+i\kappa$) and negative ($-i\kappa$) directions of the $y$-axis.

Based on principal characteristics of superlattices \cite{Bass_book_1997} we further stipulate that all characteristic dimensions $d_m$, $d_s$ and $L$ of the structure under study satisfy the long-wavelength limit, i.e. they are all much smaller than the wavelength in the corresponding layer and period ($d_m\ll \lambda$, $d_s \ll \lambda$, $L \ll \lambda$), and, thus, the multilayered system is considered to be a finely-stratified one. In view of this assumption, a standard homogenization procedure from the effective medium theory (see, Refs.~\onlinecite{Tuz_JMMM_2016, Agranovich_SolidStateCommun_1991, Eliseeva_TechPhys_2008}, and Appendix A) is applied in order to derive averaged expressions for effective constitutive parameters of the superlattice. In this way, the given finely-stratified multilayered system is approximately represented as a uniform \textit{gyroelectromagnetic} medium, whose  optical axis is directed along the structure periodicity which coincides with the direction of the external static magnetic field $\vec M$. Therefore, the resulting composite medium is a half-space that is characterized by the tensors of relative effective permeability $\hat\mu_{eff}$ and relative effective permittivity $\hat\varepsilon_{eff}$, which expressions derived via underlying constitutive parameters of magnetic ($\hat \mu_m$, $\varepsilon_m$) and semiconductor ($\mu_s$, $\hat \varepsilon_s$) layers one can find in Appendix B.

For further reference, the dispersion curves of the tensors components of relative effective permeability $\hat\mu_{eff}$ and relative effective permittivity $\hat\varepsilon_{eff}$ of the composite medium calculated according to formulas given in Appendices A and B are presented in Fig.~\ref{fig:fig_EpsMu}. Based on typical constitutive parameters which are inherent to available materials (here we follow the results of Ref.~\onlinecite{Wu_JPhysCondensMatter_2007} where a magnetic-semiconductor composite in the form of a barium-cobalt/doped-silicon superlattice is considered) we perform our calculations for the microwave band, albeit all results can be easily extrapolated to other parts of spectrum. The characteristic resonant frequencies of the underlying magnetic and semiconductor materials of the superlattice appears to be different but rather closely spaced within the same frequency band \cite{Wu_JPhysCondensMatter_2007}. Since in this paper we study only characteristics of the eigenwaves propagation, the losses in the underlying materials are neglected, i.e. the structure is considered to be \textit{non-dissipative}. Note, in the polar geometry under study the next relations between the effective tensors components hold: $\mu_{xx}=\mu_{zz}$, $\mu_{xz}=-\mu_{zx}$, $\varepsilon_{xx}=\varepsilon_{zz}$, and $\varepsilon_{xz}=-\varepsilon_{zx}$.

\begin{figure}
\centerline{\includegraphics[width=9.0cm]{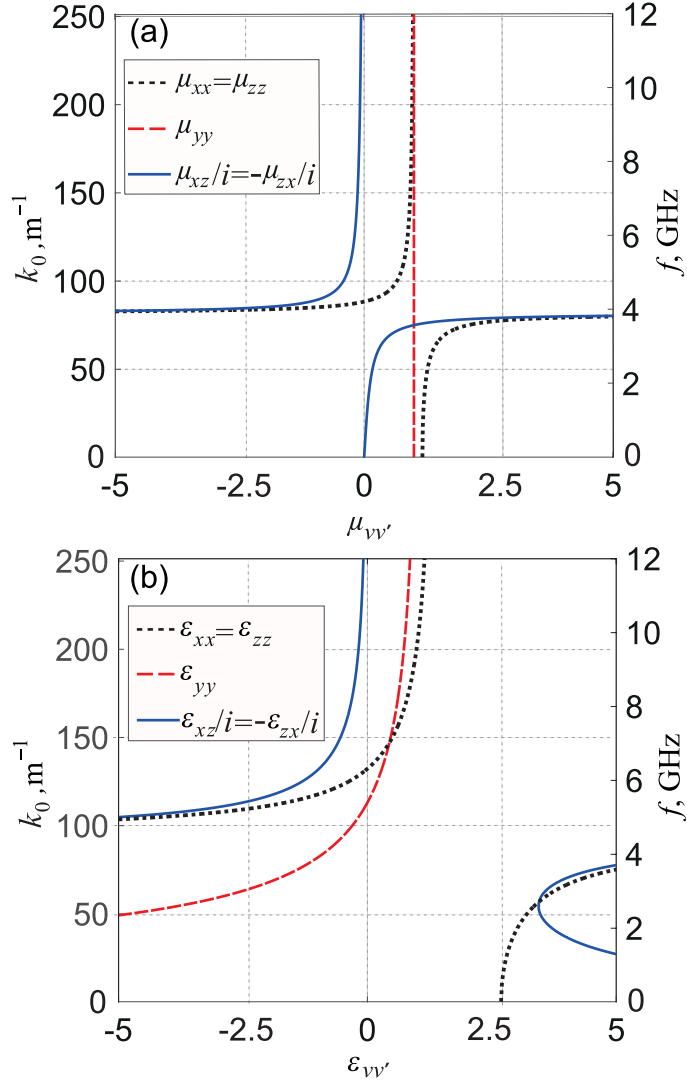}}
\caption{(Color online) Dispersion curves of the tensors components of (a) relative effective permeability $\hat\mu_{eff}$ and (b) relative effective permittivity $\hat\varepsilon_{eff}$ of the composite medium. For the magnetic constitutive layers, under saturation magnetization of 2000~G, parameters are: $f_0=\omega_0/2\pi=3.9$~GHz, $f_m=\omega_m/2\pi=8.2$~GHz, $b=0$, $\varepsilon_m=5.5$; for the semiconductor constitutive layers, parameters are: $f_p=\omega_p/2\pi=5.5$~GHz, $f_c=\omega_c/2\pi=4.5$~GHz, $\nu=0$, $\varepsilon_l=1.0$, $\mu_s=1.0$. The superlattice's geometric factors are: $\delta_m =d_m/L=0.081$ and $\delta_s=d_s/L=0.919$.}
\label{fig:fig_EpsMu}
\end{figure}

In order to derive a solution for both bulk and surface polaritons we elaborate the approach obtained in Ref.~\onlinecite{Burstein_PhysRevB_1974} where dispersion relations for polaritons in a \textit{gyroelectric} medium are derived. In the present study this approach is extended to the case of a \textit{gyroelectromagnetic} medium, whose relative permeability as well as relative permittivity simultaneously are tensor quantities (see Appendix~C; for a general treatment also see  Refs.~\onlinecite{borisov_OptSpectr_1994, Tsurumi_JPSJ_2007, Lakhtakia_book_2013}). 

The extended approach gives us two dispersion relations  which outline the areas of existence (continua) of bulk polaritons. They are obtained as follows:  
\begin{subequations}
\label{eq:bulklines}
\begin{align}
\label{eq:bulklinesa}
& k_x^2 = k_0^2\varsigma_{yy}=k_0^2\mu_{yy}\varepsilon_{yy},\\
\label{eq:bulklinesb}
& k_x^2 = k_0^2\left(\varsigma_{zz}-\frac{\varsigma_{xz}\varsigma_{zx}}{\varsigma_{xx}}\right) = k_0^2\mu_v\varepsilon_v\left(1-\frac{\mu_{xz}}{\mu_{xx}}\frac{\varepsilon_{xz}}{\varepsilon_{xx}}\right)^{-1}, 
\end{align}
\end{subequations}
where $\varsigma_{\nu\nu'}$ are elements of the tensor $\hat \varsigma$ which is introduced as the product of tensors $\hat \mu_{eff}$ and $\hat \varepsilon_{eff}$ made in the appropriate order (in what follows subscripts $\nu$ and $\nu'$ are substituted to iterate over corresponding indexes of the tensor quantities in Cartesian coordinates), and $\mu_v=\mu_{xx} + \mu_{xz}^2/\mu_{xx}$ and $\varepsilon_v=\varepsilon_{xx} + \varepsilon_{xz}^2/\varepsilon_{xx}$ are introduced as the effective bulk permeability and permittivity, respectively.

The dispersion equation for surface polaritons is derived in the form: 
\begin{equation}
\begin{split}
\kappa_0^2\frac{g_{xx}}{g_0}&\left\{\left(\kappa_1^2+\kappa_1\kappa_2+\kappa_2^2-\mbox{\ae}_z^2\right)+\mbox{\ae}_y^2\frac{\varsigma_{xz}}{\varsigma_{yy}}\frac{g_{xz}}{g_{xx}}\right\} + \kappa_0\left\{ \kappa_1\kappa_2(\kappa_1+\kappa_2)+\mbox{\ae}_y^2\frac{g_{xx}}{g_0}\frac{g_v}{\varsigma_{yy}}(\kappa_1+\kappa_2)\right\} \\
& + \frac{g_{xz}}{\varsigma_{xz}}\biggl\{\left(\mbox{\ae}_z^4-\mbox{\ae}_z^2\left(\kappa_1^2+\kappa_2^2\right)+\kappa_1^2\kappa_2^2\right) + \frac{\varsigma_{xz}}{\varsigma_{yy}}\frac{g_{zz}}{g_{xz}}\mbox{\ae}_y^2\left(\mbox{\ae}_z^2+\kappa_1\kappa_2\right)\biggr\}=0,
\end{split}
\label{eq:dispeq}
\end{equation}
where $\mbox{\ae}_\nu^2=k_x^2-k_0^2\varsigma_{\nu\nu}$, and two distinct substitutions $\varepsilon_{\nu\nu'} \to g_{\nu\nu'}$, $\varepsilon_0 \to g_0$ and $\mu_{\nu\nu'} \to g_{\nu\nu'}$, $\mu_0 \to g_0$ correspond to the problem resolving with respect to vectors $\vec E$ and $\vec H$, respectively (here we kindly ask the reader to compare two solution procedures described in Refs.~\onlinecite{Burstein_PhysRevB_1974} and \onlinecite{Elmzughi_PhysRevB_1995} for gyroelectric (semiconductor) and gyromagnetic (ferrite) superlattices, respectively). The remaining notations are given in Appendix~C.

Notice, in two particular cases of a medium which is characterized by either scalar permeability ($\mu_{eff}$) and tensor permittivity ($\hat\varepsilon_{eff}$), or tensor permeability ($\hat\mu_{eff}$) and scalar permittivity ($\varepsilon_{eff}$), dispersion relation \eqref{eq:dispeq} coincides with Eq.~(23) of Ref.~\onlinecite{Burstein_PhysRevB_1974} and Eq. (21) of Ref.~\onlinecite{Elmzughi_PhysRevB_1995} for gyroelectric (semiconductor) and gyromagnetic (ferrite) superlattices, respectively, that verifies the obtained solution.

Since $\mbox{\ae}_y$ and $\mbox{\ae}_z$ in Eq.~\eqref{eq:dispeq} emerge only in even powers, dispersion features of surface polaritons appear to be identical for positive and negative $k_x$ directions. This means, that their dispersion characteristics possess a reciprocal nature. 

\section{\label{sec:plasmons}III. Dispersion of Bulk and Surface Polaritons in a~Gyroelectromagnetic Medium}
\subsection{\label{sec:plasmonsa}1. Continua of Bulk Polaritons }

In this section we reveal the continua and dispersion features  of bulk polaritons, whose appearance in the $k_0-k_x$ plane is defined by relations in \eqref{eq:bulklines}. In accordance with the problem geometry (see, Fig.~\ref{fig:fig_Struct}), Eq. \eqref{eq:bulklinesa} corresponds to the electromagnetic field with components $\{E_x,H_y,E_z\}$, whereas Eq. \eqref{eq:bulklinesb} corresponds to the field with components $\{H_x,E_y,H_z\}$. One can see that in the former case the magnetic field vector is parallel to the external magnetic field $\vec M$, which results in absence of its interaction with the magnetic system. Since $\mu_{yy}$ is a constant quantity ($\mu_{yy}=1$) within the whole frequency band of interest, continua of bulk polaritons which is defined by Eq.~\eqref{eq:bulklinesa} depends only on the dispersion characteristics of $\varepsilon_{yy}$. Contrariwise, Eq.~(\ref{eq:bulklinesb}) outlines areas of existence of bulk polaritons, whose dispersion features are influenced by the external magnetic field, and they depend on the resonant characteristics of both effective bulk permeability $\mu_v$ and effective bulk permittivity $\varepsilon_v$. As is convenient in the plasma physics \cite{ginzburg_book_1961}, we further distinguish these two different sorts of waves as \textit{ordinary} and \textit{extraordinary} bulk polaritons, respectively. Moreover, in view of the forms of Eqs.~\eqref{eq:bulklinesa} and \eqref{eq:bulklinesb} one can conclude that the dispersion features of ordinary bulk polaritons should appear to be quite trivial, while those of extraordinary bulk polaritons can possess some peculiarities which require  a special consideration. 
 
\begin{figure*}
\centerline{\includegraphics[width=13.5cm]{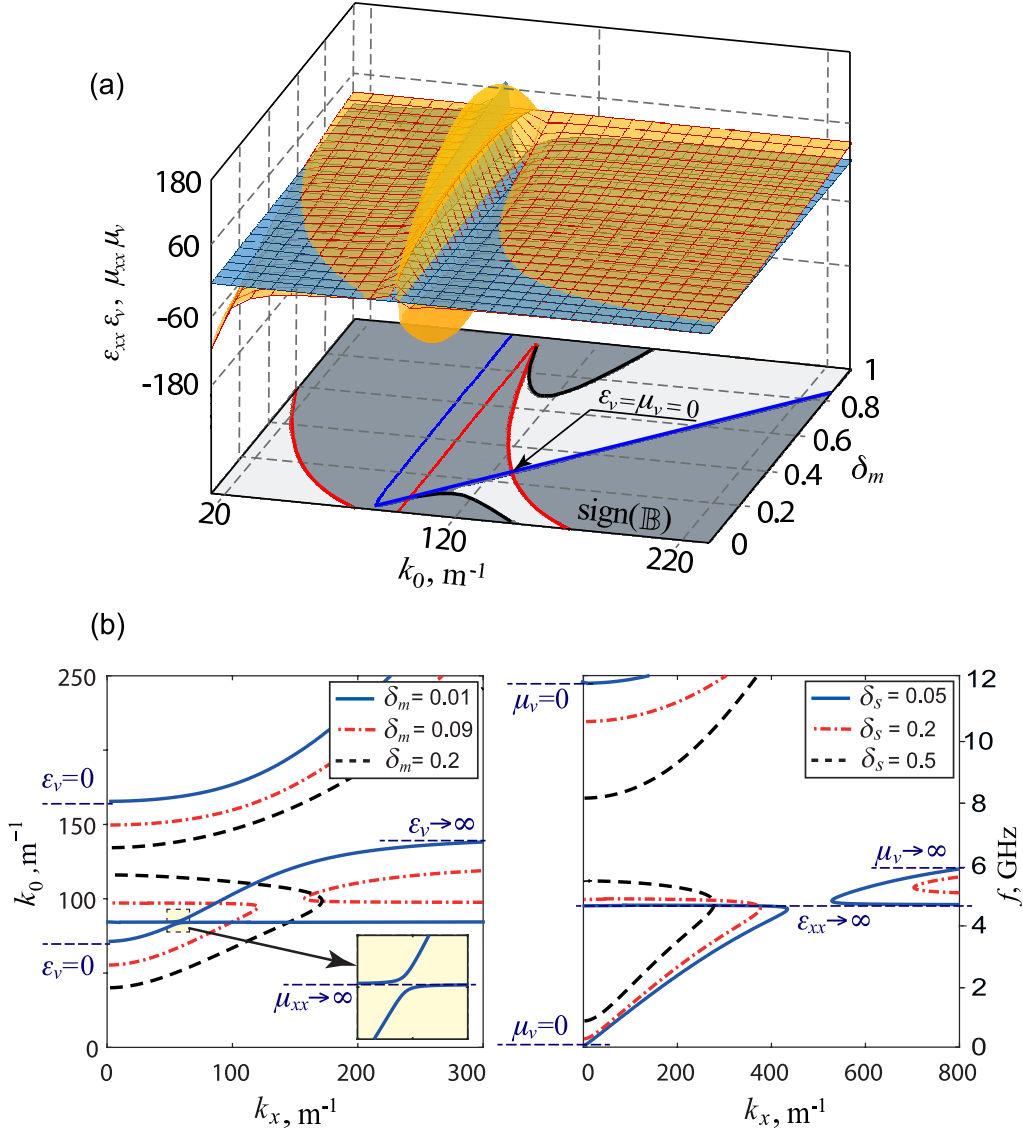}}
\caption{(Color online) (a) Characterization of areas of existence (continua) and nonexistence of extraordinary bulk polaritons. Two surfaces at the top depict numeric values of  effective bulk permeability $\mu_v$ (blue surface) and  effective bulk permittivity $\varepsilon_v$ (orange surface). Filled contours at the bottom draw the areas where $\mathbb{B}$ is either positive (light gray areas) or negative (dark gray areas). The red and blue curves outline the areas of sign changing of $\mu_v$ and $\varepsilon_v$, respectively. (b) A set of dispersion curves which outline the continua of extraordinary bulk polaritons for different filling factor $\delta$. All structure constitutive parameters are as in Fig.~\ref{fig:fig_EpsMu}.}
\label{fig:fig_Bulk}
\end{figure*}

First of all, extraordinary bulk polaritons can exist only if the next condition holds:
\begin{equation}
\label{eq:exbulkcond}
\mathbb{B}\equiv\mu_v\varepsilon_v\left(1-\frac{\mu_{xz}}{\mu_{xx}}\frac{\varepsilon_{xz}}{\varepsilon_{xx}}\right)^{-1}>0.    
\end{equation}
It immediately gives us four following combinations of conditions: 
\begin{subequations}
\label{eq:bulkpolaritons}
\begin{align}
\label{eq:bulkpolaritonsA}
& \mu_v>0,~~~\varepsilon_v>0,~~~ \mu_{xz}\varepsilon_{xz}/\mu_{xx}\varepsilon_{xx}<1, \\
\label{eq:bulkpolaritonsB}
& \mu_v<0,~~~\varepsilon_v>0,~~~ \mu_{xz}\varepsilon_{xz}/\mu_{xx}\varepsilon_{xx}>1, \\
\label{eq:bulkpolaritonsC}
& \mu_v>0,~~~\varepsilon_v<0,~~~ \mu_{xz}\varepsilon_{xz}/\mu_{xx}\varepsilon_{xx}>1, \\
\label{eq:bulkpolaritonsD}
& \mu_v<0,~~~\varepsilon_v<0,~~~ \mu_{xz}\varepsilon_{xz}/\mu_{xx}\varepsilon_{xx}<1,
\end{align}
\end{subequations}
the fulfillment of which guarantees the existence of extraordinary bulk polaritons. From \eqref{eq:bulkpolaritons} one can conclude that the presence of such combinations of conditions significantly extend the capabilities of existence of extraordinary bulk polaritons in the combined magnetic-semiconductor structure as compared to characteristics of conventional either magnetic or semiconductor gyrotropic medium. Note, for a non-dissipative medium the diagonal components ($\mu_{xx}$ and $\varepsilon_{xx}$) of the relative effective constitutive tensors are purely real quantities, whereas their off-diagonal components ($\mu_{xz}$ and $\varepsilon_{xz}$) are purely imaginary ones.

Therefore, in order to accurately identify areas of existence (passbands) and nonexistence (stopbands) of extraordinary bulk polaritons a multiparameter problem has been solved. As parameters of this problem the layers thicknesses and characteristic resonant frequencies of the magnetic and semiconductor layers forming the superlattice (which, in fact, depend on the physical properties of the underlying materials and the static magnetic field strength) are taken into account. In our calculations both the characteristic resonant frequencies of the underlying materials and the superlattice's period are chosen and fixed, and then the filling factor $\delta$ ($\delta_m = d_m/L$, $\delta_s = d_s/L$, $\delta_m + \delta_s = 1$) for each frequency value within the band of interest is varied. As resulting functions effective bulk permeability $\mu_v$, effective bulk permittivity $\varepsilon_v$, and  the sign of $\mathbb{B}$ in Eq. \eqref{eq:exbulkcond} are selected and plotted in Fig.~\ref{fig:fig_Bulk}(a).

From this figure one can conclude that for a particular filling factor $\delta$ there are two isolated areas of existence of extraordinary bulk polaritons. These passbands are outlined in Fig.~\ref{fig:fig_Bulk}(a) by the red and blue curves that express some combinations of the structure parameters at which the corresponding multiplier $\mu_v$ or $\varepsilon_v$ of the numerator of Eq.~\eqref{eq:bulklinesb} changes its sign. The state $\mu_v = \varepsilon_v = 0$ is considered as a crossing point between these two parameters combinations \cite{Tuz_JMMM_2016, Tuz_PIERB_2012, Tuz_JOpt_2015, Tuz_Springer_2016} (in literature this effect is also known as the mode crossing/anti-crossing effect \cite{Mertens_RadioSci_1996,Tuz_superlattice_2017}). In the bottom part of Fig.~\ref{fig:fig_Bulk}(a) this state is distinguished by an arrow. Remarkable, a passage across this state when varying the filling factor $\delta$ results in manifestation of some extremum in dispersion characteristics of extraordinary bulk polaritons. Indeed, on the appropriate side of this extremum the passbands outlines are dependent on the resonant characteristic either effective bulk permeability $\mu_v$ or effective bulk permittivity $\varepsilon_v$. 

In order to express this peculiarity more clearly, a set of dispersion curves which outline the passbands of extraordinary bulk polaritons in the $k_0-k_x$ plane is presented in Fig.~\ref{fig:fig_Bulk}(b) for a different filling factor $\delta$. Thus, in the left figure for all present values of $\delta_m$ the upper passband exists when condition \eqref{eq:bulkpolaritonsA} holds, whereas the bottom passband exists when either condition \eqref{eq:bulkpolaritonsA} or \eqref{eq:bulkpolaritonsB} holds. The upper and bottom passbands are bounded below by the line at which $\varepsilon_v = 0$ and the bottom passband is bounded above by the asymptotic line where $\varepsilon_v \to \infty$. For the right figure for all present values of $\delta_s$ the upper passband exists when condition \eqref{eq:bulkpolaritonsA} holds, whereas the bottom passband exists when either condition \eqref{eq:bulkpolaritonsA} or \eqref{eq:bulkpolaritonsC} holds. The corresponding outlines are at the lines at which $\mu_v = 0$ and $\mu_v \to \infty$, respectively (in order not to overload the drawings in these figures all asymptotic lines are designated only for the blue solid curves). 

Whereas the width and position of the bottom passbands of extraordinary bulk polaritons are defined by the corresponding resonant frequencies of effective bulk permeability $\mu_v$ and effective bulk permittivity $\varepsilon_v$, which, in fact, are multipliers of the numerator of Eq.~\eqref{eq:bulklinesb}, its denominator originates a singularity at the asymptotic line where $1 - \mu_{xz} \varepsilon_{xz} / \mu_{xx} \varepsilon_{xx} \to 0$. Obviously, it corresponds to the cases when either $\mu_{xx} \to \infty$ or $\varepsilon_{xx} \to \infty$. This asymptotic line splits the bottom passbands on two separated sub-passbands which is distinguished in the left inset of Fig.~\ref{fig:fig_Bulk}(b). As the corresponding filling factor $\delta_m$ or $\delta_s$ rises these two separated sub-passbands transform into closed continua. This continua exist when condition \eqref{eq:bulkpolaritonsA} holds, in particular, in the case when the next combination is met: $\mu_v>0, \varepsilon_v>0$, and $|\mu_{xz}\varepsilon_{xz}|>|\mu_{xx}\varepsilon_{xx}|$. It should be mentioned, for the considered structure configuration, there are not any continua for which condition \eqref{eq:bulkpolaritonsD} holds.

\begin{figure}
\centerline{\includegraphics[width=10.0cm]{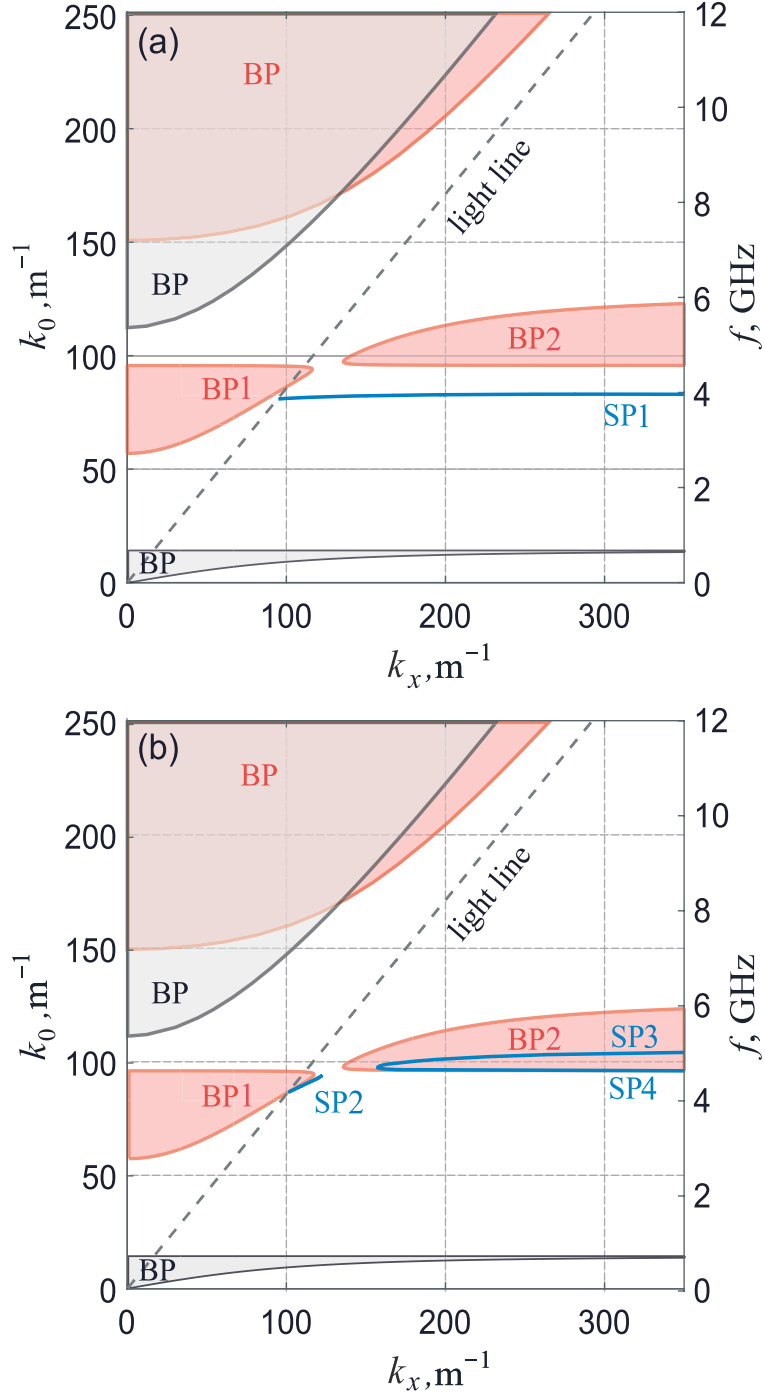}}
\caption{(Color online) Areas of existence of both ordinary (gray areas) and extraordinary (red areas) bulk polaritons (`BP') and the dispersion curves (blue solid lines) of the surface polaritons (`SP') in the case of (a) EH-predominant subsystem and (b) HE-predominant subsystem. All structure constitutive parameters are as in Fig.~\ref{fig:fig_EpsMu}; $\delta_m = 0.081$; $\delta_s = 0.919$.}
\label{fig:fig_Surf}
\end{figure}

The passbands of both ordinary and extraordinary bulk polaritons are also depicted on the same $k_0-k_x$ plane in Fig.~\ref{fig:fig_Surf} for a particular filling factor $\delta$, and they are distinguished from each other by abbreviation `BP' and different colors. Thus, the areas colored in gray and red are related to the passbands of ordinary and extraordinary bulk polaritons, respectively.

One can conclude that ordinary bulk polaritons demonstrate typical behaviors having two passbands separated by a stopband. The bottom passband starts from zero frequency and it is bounded above by the asymptotic line where $\varepsilon_{yy}\to\infty$, while the upper passband is bounded laterally by the light line and its lower limit is restricted by the line at which $\varepsilon_{yy}=0$. 

For the extraordinary bulk polaritons the upper passband has a typical form and its lower limit is restricted by the line at which $\varepsilon_v=0$. At the same time, the bottom passband is limited by the lines at which $\varepsilon_v = 0$ and $\varepsilon_v \to \infty$, and splits into two separated sub-passbands by the asymptotic line where $\mu_{xx} \to \infty$. These sub-passbands are denoted in Fig.~\ref{fig:fig_Surf} as `BP1' and `BP2'. Remarkably, the different branches of the dispersion curves that outline these sub-passbands can manifest normal as well as anomalous dispersion.     

\subsection{\label{sec:plasmonsb}2. Dispersion Curves of Surface Polaritons}

In order to calculate the dispersion curves of surface polaritons, dispersion equation \eqref{eq:dispeq} is solved numerically. In general, it allows two considerations providing a particular substitution into Eq.~\eqref{eq:dispeq} of $\varepsilon_0 \to g_0$, $\varepsilon_{\nu\nu'} \to g_{\nu\nu'}$ or $\mu_0 \to g_0$, $\mu_{\nu\nu'} \to g_{\nu\nu'}$ which implies the problem resolving with respect to vector $\vec H$ or $\vec E$ (i.e. we substitute $\vec H\to \vec P$ or $\vec E\to \vec P$ into initial wave equation \eqref{eq:waveeq} (here we kindly ask the reader to compare the solution procedures given in Refs.~\onlinecite{Burstein_PhysRevB_1974} and \onlinecite{Elmzughi_PhysRevB_1995} for the gyroelectric and gyromagnetic superlattices, respectively). In fact, these two considerations result in different relations between the magnitudes of the transverse and longitudinal electric and magnetic fields components. It is specific for the \textit{hybrid} waves \cite{Mertens_RadioSci_1996, Ivanov_NATO_2010}, since in our case the resulting electromagnetic field has all six nonzero components. According to the hybrid waves taxonomy, in what follows we distinguish between these two considerations with the terms \textit{EH-predominant subsystem} and \textit{HE-predominant subsystem}, respectively.

Therefore, we classify the hybrid waves as the modes that have either EH-type or HE-type depending on the magnitudes ratio between the longitudinal electric and magnetic fields components \cite{Ivanov_NATO_2010}. Thus, it is supposed that for the large values of $k_x$ the wave has the EH-type if $E_x > H_x$ and the HE-type if $H_x > E_x$. Note, the wave type can be different within the same dispersion curve for different values of $k_x$. At small values of $k_x$ the wave can change its hybrid type repeatedly, whereas for the large values of $k_x$ it remains to have the same type along the dispersion curve \cite{Ivanov_NATO_2010} and it is suitable for the modes classification.

For each predominant subsystem, dispersion equation \eqref{eq:dispeq} of the surface polaritons has four roots. As was already mentioned, from these four roots those two must be selected which ensure the wave attenuation as it propagates, that imposes restrictions on the values of $\kappa_1$ and $\kappa_2$ derived from Eq.~\eqref{eq:kappa11}. Depending upon the position in the $k_0-k_x$ plane, in the \textit{non-dissipative} system the following combinations between $\kappa_1$ and $\kappa_2$ may arise \cite{Burstein_PhysRevB_1974}: (i) both roots are real and positive (bonafide surface modes); (ii) one root is real and the other is pure imaginary, or vice versa (pseudosurface modes); (iii) both roots are complex in which case they are conjugate (generalized surface modes); (iv) both roots are pure imaginary (the propagation is forbidden). 

In our study we are interested only in the bonafide surface modes, therefore, the root branches of Eq.~\eqref{eq:dispeq} are properly selected and plotted in Fig.~\ref{fig:fig_Surf} for two different predominant subsystems. Thus, the blue solid curve in Fig.~\ref{fig:fig_Surf}(a) with abbreviation `SP1' is related to the EH-predominant subsystem, whereas three blue solid curves in Fig.~\ref{fig:fig_Surf}(b) with abbreviations `SP2'-`SP4' are related to the HE-predominant subsystem, and correspondingly for the large values of $k_x$ the surface wave whose dispersion conditions are defined by the curve `SP1' possesses the EH-type, whereas those defined by the dispersion curves `SP2'-`SP4' possess the HE-type.

In order to identify the area of existence and asymptotic lines of  surface polaritons the \textit{magnetostatic} (nonretarded) limit is considered. This limit corresponds to the condition $k_x\gg k_0$ (which is mathematically equivalent to $c \to \infty $). Then taking into account that $\mbox{\ae}_\nu=\kappa_0=k_x$, from Eq.~\eqref{eq:kappa11} we have: 
\begin{subequations}
\label{eq:nonretarded}
\begin{align}
& \kappa_1^2 = k_x^2\frac{\varsigma_{xx}}{\varsigma_{yy}}, \\
& \kappa_2^2=k_x^2,
\end{align}
\end{subequations}
and from \eqref{eq:dispeq} the band of existence of surface polaritons can be found as the next:
\begin{equation}
\label{eq:dispnonretarded}
\mathbb{S}\equiv\eta\left[\left(g_0+g_{xx}\right)\chi+\left(g_0+g_{v}\right)g_{xx}\right](\varsigma_{xz}g_{xz})^{-1} > -1,
\end{equation}
where $\eta = 1 + (\varsigma_{xx}/\varsigma_{yy})^{1/2}$ and $\chi = (\varsigma_{xx}\varsigma_{yy})^{1/2}$, and the condition $\mathbb{S}=-1$ gives us the required asymptotic limits. 

In the non-dissipative system inequality \eqref{eq:dispnonretarded} can be met only in the field of real numbers which inevitably entails that $\varsigma_{xx}$ and $\varsigma_{yy}$ must have the same sign (i.e. $\eta$ and $\chi$ must be real numbers). As the resonant frequencies of the magnetic and semiconductor materials of the superlattice under study appear to be closely spaced within the same frequency band, inequality \eqref{eq:dispnonretarded} can be satisfied. It is important to note that in our case $\mu_{yy}$ is a positive constant quantity, therefore, the domain of existence (real or complex) of the values $\eta$ and $\chi$ depends on the resonant frequency of $\varepsilon_{yy}$ only. In fact, the branches of surface polaritons  arise in the bands where both $\varsigma_{xx}$ and $\varsigma_{yy}$ are negative quantities. 

Furthermore, it is clear that two particular substitutions into inequality \eqref{eq:dispnonretarded} of the corresponding values related to the EH- and HE-predominant subsystems should obviously give the different dispersion features of surface polaritons. Indeed, for the EH-predominant subsystem inequality \eqref{eq:dispnonretarded} is satisfied in a single frequency band and only one dispersion curve arises. One can see in Fig.~\ref{fig:fig_Surf}(a) that the dispersion curve `SP1' manifests typical behaviors existing in the band where propagation of bulk polaritons is forbidden. It possesses a normal dispersion feature and starts from the light line, rises just to the right of the light line, flattens out, and then approaches the asymptotic line where $\mathbb{S}=-1$.  

In contrast with the characteristics of the EH-predominant subsystem, in the HE-predominant subsystem inequality \eqref{eq:dispnonretarded} is satisfied in two separated bands. It results in the fact that the branch of surface polaritons appears to be discontinuous which is depicted in Fig.~\ref{fig:fig_Surf}(b). Thus, the curve `SP2' arises from the light line and then ends abruptly. The curve `SP3' appears as some prolongation of the curve `SP2' which manifests normal dispersion and approaches the upper limit where $\mathbb{S}=-1$. At the same time, at the bottom part this curve continues into another branch marked `SP4' exhibiting an anomalous dispersion. We should note, earlier it was reported in \cite{tarkhanyan_PIERB_2012} that such a form of the dispersion curve is also inherent to surface magnon-polaritons in an enantiomeric antiferromagnetic (bi-anisotropic) structure.

The distinguishing feature of the structure under study is that the branch `SP3'-`SP4' arises within the passband `BP2' of extraordinary bulk polaritons. The effect appears exactly in the band where condition (\ref{eq:bulkpolaritonsB}) for extraordinary bulk polaritons holds. At the same time, from inequality \eqref{eq:dispnonretarded} it follows that for the EH-predominant subsystem surface polaritons can propagate when $\varepsilon_v$ is a negative quantity, whereas for the HE-predominant subsystem it is when $\mu_v$ is a negative quantity. As it was already discussed the passband splitting of extraordinary bulk polaritons for the corresponding filling factor $\delta_m$ appears at the line where $\mu_{xx}\to \infty$ that gives $\mu_v\to 0$. Therefore, above this line $\mu_v$ becomes to be negative quantity which allows the propagation of surface polaritons.

\section{\label{sec:conclusion}V. Conclusions}

In the present paper the effect of coexistence of bulk and surface polaritons in a magnetic-semiconductor superlattice which is influenced by an external static magnetic field in the polar geometry is revealed and discussed. 

In view of superior properties of such gyroelectromagnetic superlattice, whose magnetic and dielectric resonant frequencies of the underlying materials appear to be closely spaced within the same frequency band, two remarkable results are distinguished. First, the continua of bulk polaritons can split into two separated passbands, and, second, in one of these passbands  surface polaritons can propagate. It has been shown, that such coexistence of bulk and surface polaritons within the same frequency band and wavevector space can be obtained by providing an appropriate choice of the superlattice constitutive parameters and the structure filling factor.

Although from a theoretical viewpoint (with respect to eigenwaves) the existence of the discussed effect in the polariton spectra is not in doubt, the effect of losses and stability of waves propagation in practical systems are still required a particular consideration since the magnetic dumping and charge carrier collisions can lead to the dispersion curves smoothing, decreasing in the amplitude of the observed phenomena, disappearance
of resonances and, consequently, changes in the properties of the existing modes. 
 
We expect that the effect of coexistence of bulk and surface polaritons can be also found in chiral (bi-isotropic) media, and it must inevitably occur at the interface of a general class of bi-anisotropic media. Moreover, it can give great advantages when providing excitation of surface polaritons via their nonlinear coupling with bulk waves.

\appendix

\setcounter{equation}{0}
\numberwithin{equation}{section}
\renewcommand{\theequation}{A.\arabic{equation}}

\section{\label{sec:effmedium}Appendix A. Effective Constitutive Parameters  of a Superlattice}

In the long-wavelength limit ($d_m\ll\lambda$, $d_s\ll\lambda$, and $L\ll\lambda$), with the effective-medium approximation \cite{Agranovich_SolidStateCommun_1991, Eliseeva_TechPhys_2008}, the superlattice is treated as an anisotropic uniform medium, which can be illustrated by tensors of effective permeability $\hat\mu_{eff}$ and effective permittivity $\hat\varepsilon_{eff}$ that should be retrieved.

In a general form \cite{Tuz_JMMM_2016}, constitutive equations $\vec B = \hat\mu \vec H$ and
$\vec D = \hat\varepsilon \vec E$ for magnetic $(0<z<d_m)$ and semiconductor $(d_m<z<L)$ layers can be represented as follow:
\begin{equation}
Q^{(j)}_\nu =\sum_{\nu'} g^{(j)}_{\nu\nu'}P^{(j)}_{\nu'}, \label{eq:constit}
\end{equation}
where $\vec  Q$ is substituted for the magnetic and electric flux densities $\vec B$ and $\vec D$; $\vec P$ is substituted for the magnetic and electric field strengths $\vec H$ and $\vec E$; $g$ is substituted for permeability and permittivity $\mu$ and $\varepsilon$; the superscript $j$ is introduced to distinguish between magnetic $(m \to j)$ and semiconductor $(s \to j)$ layers, and $\nu,\nu'$ iterate over $x,y,z$.

In the chosen problem geometry, the $y$-axis is perpendicular to the interfaces between the layers within the structure, and, therefore, components $P^{(j)}_x$, $P^{(j)}_z$, and $Q^{(j)}_y$ are continuous. Thus, the particular component $P^{(j)}_y$  can be expressed  from Eq.~\eqref{eq:constit} in terms of the continuous components of the field
\begin{equation}
P^{(j)}_y= - \frac{g^{(j)}_{yx}}{g^{(j)}_{yy}}P^{(j)}_x + \frac{1}{g^{(j)}_{yy}}Q^{(j)}_y - \frac{g^{(j)}_{yz}}{g^{(j)}_{yy}}P^{(j)}_z, \label{eq:induction_1}
\end{equation}
and substituted into equations for components $Q^{(j)}_x$ and
$Q^{(j)}_z$:
\begin{equation}
\begin{split}
Q^{(j)}_x=&\left(g^{(j)}_{xx} - \frac{g^{(j)}_{xy}g^{(j)}_{yx}} {g^{(j)}_{yy}}\right) P^{(j)}_x + \frac{g^{(j)}_{xy}}{g^{(j)}_{yy}}Q^{(j)}_y + \left(g^{(j)}_{xz} - \frac{g^{(j)}_{xy}g^{(j)}_{yz}}{g^{(j)}_{yy}}\right) P^{(j)}_z, \\
Q^{(j)}_z=& \left(g^{(j)}_{zx} - \frac{g^{(j)}_{zy}g^{(j)}_{yx}}{g^{(j)}_{yy}}\right) P^{(j)}_x + \frac{g^{(j)}_{zy}}{g^{(j)}_{yy}}Q^{(j)}_y + \left(g^{(j)}_{zz} - \frac{g^{(j)}_{zy}g^{(j)}_{yz}}{g^{(j)}_{yy}}\right) P^{(j)}_z.
\end{split}
\label{eq:induction_2}
\end{equation}
Then these obtained relations (\ref{eq:induction_1}) and (\ref{eq:induction_2}) are used for the fields averaging \cite{Agranovich_SolidStateCommun_1991}.

Since in the long-wavelength limit the fields $\vec P^{(j)}$ and $\vec Q^{(j)}$ inside the layers are considered to be constant, the \textit{averaged} (Maxwell) fields $\langle\vec  Q\rangle$ and $\langle\vec P\rangle$ can be determined by the equalities
\begin{equation}
\langle \vec P\rangle=\frac{1}{L}\sum_j \vec P^{(j)} d_j, ~~~~~ \langle \vec Q\rangle=\frac{1}{L}\sum_j \vec Q^{(j)} d_j.
\label{eq:mean_1}
\end{equation}
In view of the above discussed continuity of components $P^{(j)}_x$,
$P^{(j)}_z$, and $Q^{(j)}_y$, it follows that
\begin{equation}
\langle P_x \rangle = P^{(j)}_x, ~~~\langle P_z \rangle = P^{(j)}_z,~~~ \langle Q_y\rangle= Q^{(j)}_y, \label{eq:conditions}
\end{equation}
and on the basis of Eqs.~\eqref{eq:induction_1} and \eqref{eq:induction_2}, the relations between the averaged fields components are obtained as:
\begin{equation}
\begin{split}
\langle Q_x\rangle&=\alpha_{xx}\langle P_x\rangle +
\gamma_{xy}\langle Q_y\rangle + \alpha_{xz}\langle P_z\rangle, \\
\langle P_y\rangle&=\beta_{yx}\langle P_x\rangle +
\beta_{yy}\langle Q_y\rangle + \beta_{yz}\langle P_z\rangle, \\
\langle Q_z\rangle&=\alpha_{zx}\langle P_x\rangle +
\gamma_{zy}\langle Q_y\rangle + \alpha_{zz}\langle P_z\rangle,
\end{split}
\label{eq:mean_2}
\end{equation}
where $\alpha_{\nu\nu'} = \sum_j
(g^{(j)}_{\nu\nu'}-g^{(j)}_{\nu y}g^{(j)}_{y\nu'}/g^{(j)}_{yy})\delta_j$, $\beta_{yy} = \sum_j (1/g^{(j)}_{yy})\delta_j$, $\beta_{y\nu'} =-\sum_j (g^{(j)}_{y\nu'}/g^{(j)}_{yy})\delta_j$, $\gamma_{\nu y}=\sum_j (g^{(j)}_{\nu y}/g^{(j)}_{yy})\delta_j$, $\delta_j=d_j/L$, and $\nu,\nu'$ iterate over $x,z$.

Expressing $\langle Q_y\rangle$ from the second equation in system (\ref{eq:mean_2}) and substituting it into the rest two equations, the constitutive equations for the flux densities of the effective medium $\langle\vec Q\rangle = \hat g_{eff} \langle\vec P\rangle$ can be derived, where $\hat g_{eff}$ is a tensor
\begin{equation}
\hat g_{eff}=\left( {\begin{matrix} {\tilde \alpha_{xx}} & {\tilde \gamma_{xy}} & {\tilde \alpha_{xz}} \cr {\tilde \beta_{yx}} & {\tilde \beta_{yy}} & {\tilde \beta_{yz}} \cr {\tilde \alpha_{zx}} & {\tilde \gamma_{zy}} & {\tilde \alpha_{zz}} \cr \end{matrix} } \right) \label{eq:gem}
\end{equation}
with components $\tilde \alpha_{\nu\nu'} = \alpha_{\nu\nu'}-\beta_{y\nu'}\gamma_{\nu y}/\beta_{yy}$, $\tilde \beta_{yy} = 1/\beta_{yy}$, $\tilde \beta_{y\nu'} = -\beta_{y\nu'}/\beta_{yy}$, and $\tilde \gamma_{\nu y} = \gamma_{\nu y}/\beta_{yy}$.

For the geometry under consideration we have $\tilde\gamma_{xy} = \tilde \gamma_{zy}= \tilde \beta_{yx}= \tilde \beta_{yz} = 0$. The other  tensors components are
\begin{equation}
\begin{split}
&\tilde \alpha_{xx} =g^{(f)}_{xx}\delta_f+g^{(s)}_{xx}\delta_s,~~~~\tilde \alpha_{zz}=g^{(f)}_{zz}\delta_f+g^{(s)}_{zz}\delta_s, \\ & \tilde \alpha_{xz} = -\tilde \alpha_{zx} = g^{(f)}_{zx}\delta_f+g^{(s)}_{zx}\delta_s,~~~~\tilde \beta_{yy} = g^{(f)}_{yy}g^{(s)}_{yy}\tau,
\end{split}
\label{eq:gemz}
\end{equation}
where  $\tau = (g^{(s)}_{yy}\delta_f+g^{(f)}_{yy}\delta_s)^{-1}$, and $\hat g^{(f)}$, $\hat g^{(s)}$ are the tensors of the underlying constitutive parameters of magnetic and semiconductor layers, respectively.

\renewcommand{\theequation}{B.\arabic{equation}}

\section{\label{sec:constpar}Appendix B. Constitutive Parameters of Magnetic and Semiconductor Layers}
\setcounter{equation}{0}

The expressions for tensors components of the underlying constitutive parameters of magnetic $\hat g^{(f)}$ and semiconductor $\hat g^{(s)}$ layers can be written in the form

\begin{equation}
\hat g^{(j)}=\left( {\begin{matrix}
   {g_1} & {0} & {ig_2} \cr
   {0} & {g_3} & {0} \cr
   {-ig_2} & {0} & {g_1} \cr
\end{matrix}
} \right). \label{eq:gfs}
\end{equation}

For magnetic layers \cite{Collin_book_1992} the components of tensor $\hat g^{(f)}$ are $g_1=1+\chi' + i\chi''$, $\quad g_2=\Omega'+i\Omega''$, $ g_3=1$, and $\quad\chi'=\omega_0\omega_m[\omega^2_0-\omega^2(1-b^2)]D^{-1}$, $\chi''=\omega\omega_m b[\omega^2_0+\omega^2(1+b^2)]D^{-1}$, $\quad\Omega'=\omega\omega_m[\omega^2_0-\omega^2(1+b^2)]D^{-1}$, $\Omega''=2\omega^2\omega_0\omega_m bD^{-1}$, $\quad D=[\omega^2_0-\omega^2(1+b^2)]^2+4\omega^2_0\omega^2 b^2$, where $\omega_0$ is the Larmor frequency and $b$ is a dimensionless damping constant.

For semiconductor layers \cite{Bass_book_1997} the components of tensor $\hat g^{(s)}$ are $g_1=\varepsilon_l[1-\omega_p^2(\omega+i\nu)[\omega((\omega+i\nu)^2-\omega_c^2)]^{-1}]$, $g_2=\varepsilon_l\omega_p^2\omega_c[\omega((\omega+i\nu)^2-\omega_c^2)]^{-1}$, $g_3=\varepsilon_l\left[{1-\omega_p^2[\omega(\omega+i\nu)]^{-1}}\right]$, where $\varepsilon_l$ is the part of permittivity attributed to the lattice, $\omega_p$ is the plasma frequency, $\omega_c$ is the cyclotron frequency and $\nu$ is the electron collision frequency in plasma.

Permittivity $\varepsilon_m$ of the magnetic layers as well as permeability $\mu_s$ of the semiconductor layers are scalar quantities.

\renewcommand{\theequation}{C.\arabic{equation}}

\section{\label{sec:dispersion}Appendix C. Solution for Bulk and Surface Polaritons}
\setcounter{equation}{0}

In a general form \cite{Tuz_JMMM_2016}, the electric and magnetic field vectors $\vec E$ and $\vec H$ used here are represented as
\begin{equation}
\vec P^{(j)}= \vec p^{(j)}\exp\left(ik_x x\right)\exp\left(\mp\kappa_j y\right), 
\label{eq:inc}
\end{equation}
where a time factor $\exp\left(-i \omega t\right)$ is also supposed and omitted, and sign `$-$' is related to the fields in the upper medium ($y>0$, $j=0$), whereas sign `$+$' is related to the fields in the composite medium ($y<0$, $j=1$).

From a pair of the curl Maxwell's equations $\nabla\times\vec E = i k_0 \vec B$ and $\nabla\times\vec H = -i k_0 \vec D$, where $k_0=\omega/c$ is the free space wavenumber, in a standard way we derive the following equation for the macroscopic field:
\begin{equation}
\nabla \times \nabla \times\vec P^{(j)} - k_0^2 \hat \varsigma^{(j)} \vec P^{(j)} = 0.
\label{eq:waveeq}
\end{equation}

For the upper medium ($j=0$), direct substitution of expression (\ref{eq:inc}) with $\vec P^{(0)}$ and corresponding constitutive parameters ($\varsigma^{(0)}\equiv\varsigma_0 \hat I=\varepsilon_0\mu_0\hat I$, where $\hat I$ is the identity tensor) into Eq.~\eqref{eq:waveeq} gives us the relation with respect to $\kappa_0$:
\begin{equation}
\kappa_0^2 = k_x^2-k_0^2\varepsilon_0\mu_0. \label{eq:kappa0}
\end{equation}

For the composite medium ($j=1$), substitution of (\ref{eq:inc}) with $\vec P^{(1)}$ and $\hat \varsigma^{(1)}\equiv\{\varsigma_{\nu\nu'}\}$ into Eq.~\eqref{eq:waveeq} with subsequent elimination of $P^{(1)}_y$ yields us the following system of two homogeneous algebraic equations for the rest two components of $\vec P^{(1)}$:
\begin{subequations}
\label{eq:systemAB}
\begin{align}
  \label{eq:systemABa}
  A_{xz}P^{(1)}_x + B_{xz}P^{(1)}_z&=0, \\
  \label{eq:systemABb}
  B_{zx}P^{(1)}_x + A_{zx}P^{(1)}_z&=0,
\end{align}
\end{subequations}
where coefficients $A_{xz}= (\kappa^2/\mbox{\ae}_y^2)k_x^2-\kappa^2-k_0^2\varsigma_{xx}$ and $A_{zx}= k_x^2-\kappa^2-k_0^2\varsigma_{zz}$ are functions of $\kappa$; $B_{\nu\nu'}= -k_0^2\varsigma_{\nu\nu'}$.

In order to find a nontrivial solution of system \eqref{eq:systemAB}, we set its determinant of coefficients to zero. After disclosure of the determinant, we obtain a biquadratic equation with respect to $\kappa$
\begin{equation}
\varsigma_{yy}\kappa^4+a\kappa^2+b=0, 
\label{eq:kappa1}
\end{equation}
where $a =-\mbox{\ae}_y^2\varsigma_{xx}-\mbox{\ae}_z^2\varsigma_{yy}$, $b =\mbox{\ae}_y^2(\mbox{\ae}_z^2\varsigma_{xx} +k_0^2\varsigma_{xz}\varsigma_{zx})$, and whose solution is
\begin{equation}
\kappa^2 =  k_x^2\left(\frac{\varsigma_{xx}+\varsigma_{yy}}{2\varsigma_{yy}}\right)-k_0^2\varsigma_{xx} \pm \left[k_x^4\left(\frac{\varsigma_{xx}-\varsigma_{yy}}{2\varsigma_{yy}}\right)^2-\frac{k_0^2\mbox{\ae}_y^2\varsigma_{xz}\varsigma_{zx}}{\varsigma_{yy}}\right]^{1/2}.
\label{eq:kappa11}
\end{equation}
The dispersion equations for bulk polaritons can be determined from Eq.~\eqref{eq:kappa1} by putting $\kappa = 0$ inside it.

In order to find the dispersion law of surface polaritons from four roots of Eq.~\eqref{eq:kappa1} those must be selected which satisfy the physical conditions, namely, wave attenuation as it propagates, that imposes restrictions on the values of $\kappa$, whose real parts must be positive quantities. In general, two such roots are required to satisfy the electromagnetic boundary conditions at the surface of the composite medium. We define these roots as $\kappa_1$ and $\kappa_2$, and then following Ref.~\onlinecite{Burstein_PhysRevB_1974} introduce the amplitudes $K_w$ ($w=1,2$) in the form:
\begin{subequations}
\label{eq:coeffKi}
\begin{align}
\label{eq:coeffKia}
P_x^{(1)}(\kappa_w) &= K_w A_{zx}(\kappa_w),  \\
\label{eq:coeffKib}
P_y^{(1)}(\kappa_w) &= K_w C(\kappa_w), \\
\label{eq:coeffKic}
P_z^{(1)}(\kappa_w) &= -K_w B_{zx}(\kappa_w),
\end{align}
\end{subequations}
where $C(\kappa_w)=-i(k_x\kappa_w/\mbox{\ae}_y^2)A_{zx}(\kappa_w)$, and these amplitudes $K_w$ need to be determined from the boundary conditions.

Taking into consideration that two appropriate roots $\kappa_1$ and $\kappa_2$ of Eq.~\eqref{eq:kappa1} are properly selected, the components of field $\vec P^{(1)}$ can be rewritten as the linear superposition of two terms with respect to these roots:
\begin{subequations}
\label{eq:inckappa}
\begin{align}
\label{eq:inckappaa}
P_x^{(1)}&= \sum_{w=1,2}K_wA_{zx}(\kappa_w)\exp(\kappa_w y),  \\
\label{eq:inckappab}
P_y^{(1)}&= \sum_{w=1,2}K_wC(\kappa_w)\exp(\kappa_w y),  \\
\label{eq:inckappac}
P_z^{(1)}&= \sum_{w=1,2}K_wB_{zx}(\kappa_w)\exp(\kappa_w y),  
\end{align}
\end{subequations}
where $y < 0$ and the factor $\exp\left[i (k_x x-\omega t)\right]$ is omitted.

Involving a pair of the divergent Maxwell's equations $\nabla \cdot \vec B=0$ and $\nabla \cdot \vec D=0$ in the form
\begin{equation}
\nabla \cdot \vec Q^{(j)}=\nabla \cdot \left(\hat g^{(j)} \vec
P^{(j)}\right)=0, \label{eq:divergent}
\end{equation}
where $\hat g^{(j)}$ is substituted for tensors of relative effective permeability $\hat \mu_{eff}$ and relative effective permittivity $\hat \varepsilon_{eff}$, and $\vec Q$ is substituted for the magnetic $\vec B$ and electric $\vec D$ flux densities, one can immediately derive the relations between the field components in the upper and composite media as follows:
\begin{subequations}
\label{eq:fieldcomp}
\begin{align}
\label{eq:fieldcompa}
&P^{(0)}_y = \frac{i k_x}{\kappa_0}P^{(0)}_x, \\
\label{eq:fieldcompb}
&P^{(1)}_y = -\frac{i k_x \kappa}{\mbox{\ae}_y^2}P^{(1)}_x.
\end{align}
\end{subequations}

The boundary conditions at the interface require the continuity of the tangential components of $\vec E$ and $\vec H$ and the normal components of $\vec D$ and $\vec B$, i.e. in our notations these components are $P_x$, $P_z$, and $Q_y$, respectively. Therefore, imposition of the boundary conditions together with relations \eqref{eq:fieldcomp} gives us the next set of four independent linear homogeneous algebraic equations with respect to unknown amplitudes $K_1$, $K_2$ and $P^{(0)}_x$, $P^{(0)}_z$:  
\begin{subequations}
\label{eq:finalsystem}
\begin{align} 
\label{eq:finalsystema}
P_x^{(0)} &= \sum_{w=1,2}K_wA_{zx}(\kappa_w) , \\ 
\label{eq:finalsystemb}
P_z^{(0)} &= -\sum_{w=1,2}K_wB_{zx}(\kappa_w), \\
\label{eq:finalsystemc} 
\frac{\kappa_0^2-k_x^2}{\kappa_0g_0}P^{(0)}_x &= \frac{g_{xx}}{\varrho}\left\{i k_x\sum_{w=1,2}K_w C(\kappa_w)\right.\left.-\sum_{w=1,2}\kappa_w K_w \left[A_{zx}(\kappa_w)-\frac{g_{zx}}{g_{xx}}B_{zx}(\kappa_w)\right]\right\},\\
\label{eq:finalsystemd}
\frac{\kappa_0}{g_0}P_z^{(0)} &= \frac{g_{xz}}{\varrho}\left\{i k_x\sum_{w=1,2}K_w C(\kappa_w)\right.\left.-\sum_{w=1,2}\kappa_w K_w \left[A_{zx}(\kappa_w)-\frac{g_{zz}}{g_{xz}}B_{zx}(\kappa_w)\right]\right\}, 
\end{align}
\end{subequations}
where $\varrho=g_{xx}g_v$, and $g_{\nu\nu'}$ are elements of the tensor $\hat g_{eff}$ which is substituted for the corresponding tensor of relative effective permeability $\hat \mu_{eff}$ or relative effective permittivity $\hat \varepsilon_{eff}$. The system of equations \eqref{eq:finalsystem} has a nontrivial solution only if its determinant vanishes. Applying this condition gives us the required dispersion equation for surface polaritons. 

Finally, the amplitudes $K_1$ and $K_2$ can be found by solving the set of linear homogeneous equations \eqref{eq:finalsystem}. They are:
\begin{equation}
\label{eq:amplitudes}
\begin{split}
& K_1= k_xB_{zx}(\kappa_2)(\kappa_0+\kappa_2), \\ & 
K_2= -k_xB_{zx}(\kappa_1)(\kappa_0+\kappa_1).
\end{split}
\end{equation}

Amplitudes $P^{(0)}_x$ and $P^{(0)}_z$ then follow from Eqs. \eqref{eq:finalsystema} and \eqref{eq:finalsystemb}.


\bibliography{FedFesTuz}

\end{document}